\documentclass{article}
\usepackage[dvips]{graphicx}
\long \def \blockcomment #1\endcomment{}

\title{Emergent spin}

\author{Michael Creutz \\
Brookhaven National Laboratory, \\
Upton, NY 11973, USA
\thanks{ This manuscript has been authored under contract
    number DE-AC02-98CH10886 with the U.S.~Department of Energy.
    Accordingly, the U.S. Government retains a non-exclusive,
    royalty-free license to publish or reproduce the published form of
    this contribution, or allow others to do so, for U.S.~Government
    purposes.  }
}
        
\date{August 2013}

\begin{document}

\maketitle

\begin{abstract}
Quantum mechanics and relativity in the continuum imply the well known
spin-statistics connection.  However for particles hopping on a
lattice, there is no such constraint.  If a lattice model yields a
relativistic field theory in a continuum limit, this constraint must
"emerge" for physical excitations.  We discuss a few models where a
spin-less fermion hopping on a lattice gives excitations which satisfy
the continuum Dirac equation.  This includes such well known systems
such as graphene and staggered fermions.  
\end{abstract}

\section{Introduction}

%%%%%%%%%%%%%%%%%%%%%
% Mike's header for slides in TeX
%\input epsf
%\input colordvi  

% from Frank 
\def\slashchar#1{\setbox0=\hbox{$#1$}           % set a box for #1
   \dimen0=\wd0                                 % and get its size
   \setbox1=\hbox{/} \dimen1=\wd1               % get size of /
   \ifdim\dimen0>\dimen1                        % #1 is bigger
      \rlap{\hbox to \dimen0{\hfil/\hfil}}      % so center / in box
      #1                                        % and print #1
   \else                                        % / is bigger
      \rlap{\hbox to \dimen1{\hfil$#1$\hfil}}   % so center #1
      /                                         % and print /
   \fi}                                         %

\def \nextslide{\medskip}

%%%%%%%%%%%%%%%%%%%%%%%%%%%%%%

As is well known, in quantum field theory the constraints of quantum
mechanics with special relativity give rise to the spin-statistics
connection.  In particular, fermions must have half integer spin.
However in a lattice theory, the lattice structure itself breaks
relativity.  In a lattice model one is free to formulate a model based
on spin-less fermions.  If one finds that the low energy excitations of
such a model have a relativistic spectrum, then these excitations must
carry half integer spin.  In some sense spin must ``emerge'' from the
dynamics.

Remarkably such models exist.  The most famous is based on
graphene.  The Hamiltonian for spin-less fermions hopping on a
hexagonal two dimensional lattice is easily diagonalized
\cite{CastroNeto:2009zz}, and the low energy excitations above the
half filled system do indeed mimic a relativistic spectrum.  And these
excitations satisfy the two dimensional Dirac equation, correspondingly
carrying half integer angular momentum \cite{Mecklenburg:2010ja}.

This situation is not unique.  Another example, discussed in more
detail below, is a two dimensional square lattice subjected to a
magnetic field of half a flux unit per elementary square.  This model
generalizes to more dimensions by threading the magnetic field through
all elementary plaquettes. This is a route to the well known staggered
fermion theory \cite{Kogut:1974ag,Susskind:1976jm,Sharatchandra:1981si}.

This paper reviews these models and discusses some of the interesting
relations with chiral symmetry and doubling issues.  Section
\ref{graphene} goes through the standard solution of the graphene
solution in the tight binding limit.  Section \ref{topology} discusses
the close ties between the doubling issues of lattice fermions and
topology.  Also we see how a chiral symmetry protects masses from an
additive renormalization.  Section \ref{2dmag} generalizes these ideas
to a square lattice in a magnetic field and makes the connection to
staggered fermions.  Section \ref{3dstag} extends this idea to higher
dimensions.  Section \ref{paths} discusses issues that can arrive in
going from the Hamiltonian version of staggered fermions to a
Euclidian path integral approach.  Section \ref{gauge} turns to the
introduction of gauge fields and an amusing property when the gauge
group is $SU(N)$ with $N$ even.  In Section \ref{gaugetopology} we
make some general observations on the effects of gauge field
topology on the lattice fermion spectrum.  In particular we present a
variation of the Nielsen-Ninomiya theorem\cite {Nielsen:1980rz} that
applies to all lattice actions including mass terms.  Finally there
are some brief conclusions in Section \ref{conclusions}.

\section {Graphene}
\label{graphene}

As is well known, the solution to a theory of fermions hopping on a
hexagonal lattice displays two Dirac cones.  With small excitations
around half filling, each of these cones represents a fermion
satisfying the Dirac equation.  Graphene, basically a two dimensional
hexagonal planar structure, represents a realization of this system
\cite{CastroNeto:2009zz}.

In the physical situation the electrons already have spin, so the
extra doubling of species with the two cones can be thought of as
representing ``flavor'' or ``isospin.''  However, from a theoretical
point of view one can consider spin-less fermions hopping on the
lattice, and then the excitations will formally acquire spin one-half.
In this section we review this solution.

\begin{figure}
\centering
\includegraphics[width=3in]{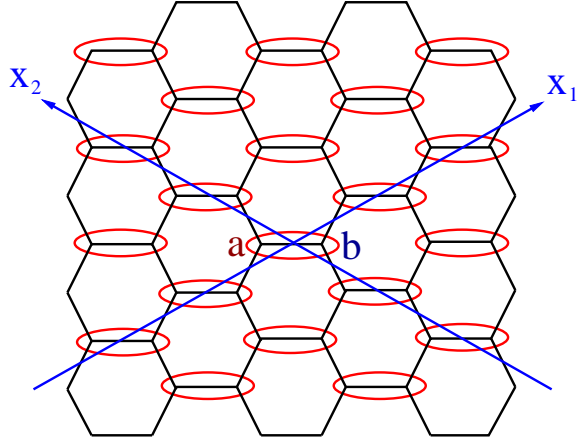}
\caption{Fermion hopping on a hexagonal lattice is nicely formulated
  in terms of $a$ and $b$ type sites labeled by non-orthogonal
  coordinates as indicated here.  Figure from
  Ref.~\cite{Creutz:2007af}.  }
\label{coordinates}
\end{figure}

To solve this problem, it is useful to use a fortuitous set of
coordinates, as sketched in Fig.~\ref{coordinates}.  Orienting the
lattice as in the figure, the sites can be considered as being of two
types.  We consider type $a$ sites on the left side of each horizontal
bond, and $b$ sites on the right.  These sites are labeled with
non-orthogonal coordinates $x_1$ and $x_2$ labeling the horizontal
bonds.  The two axes are not orthogonal, but oriented at 30 degrees
from the horizontal.  Associated with each site is a pair of
creation-annihilation operators, labeled $(a^\dagger,a)$ and
$(b^\dagger,b)$ respectively.  These satisfy the usual anti-commutation
relations
\begin{equation}
\matrix{
[a_{x_1,x_2},a^\dagger_{y_1,y_2}]_+=\delta_{x_1,y_1}\delta_{x_2,y_2}\cr
[a_{x_1,x_2},b^\dagger_{y_1,y_2}]_+=[b_{x_1,x_2},a^\dagger_{y_1,y_2}]_+=0.\cr
}
\end{equation}
With these coordinates, the nearest neighbor Hamiltonian takes the
form
\begin{eqnarray}
H=K\sum_{x_1,x_2} 
&a_{x_1,x_2}^\dagger b_{x_1,x_2}
+b_{x_1,x_2}^\dagger a_{x_1,x_2}\cr
+&a_{x_1+1,x_2}^\dagger b_{x_1,x_2}
+b_{x_1,x_2}^\dagger a_{x_1+1,x_2}\cr
+&a_{x_1,x_2}^\dagger b_{x_1,x_2+1}
+b_{x_1,x_2+1}^\dagger a_{x_1,x_2}.
\end{eqnarray}
The three terms correspond to horizontal, upward to the right, and
upward to the left bonds respectively.  Here $K$ is usually referred
to as the ``hopping'' parameter, and sets the energy scale.

To solve this system, go to momentum (reciprocal) space and define
\begin{equation}
\tilde a_{p_1,p_2}=\sum_{x_1,x_2}
\ e^{-ip_1 x_1}\ e^{-ip_2x_2}
\ a_{x_1,x_2}.
\end{equation}
These satisfy the commutation relations
\begin{equation}
[\tilde a_{p_1^\prime,p_2^\prime},\tilde a_{p_1,p_2}]_+=(2\pi)^2
\delta(p_1^\prime,p_1)\delta(p_2^\prime,p_2).
\end{equation}
Similar equations apply for the operators $b$.  Because of periodicity
in $p$, we can restrict $-\pi<p_\mu\le \pi$.  This change of variables
breaks the Hamiltonian into two by two blocks
\begin{equation}
H=K\int_{-\pi}^\pi {dp_1\over 2\pi}\  {dp_2\over 2\pi}
\ \ \pmatrix{\tilde a^\dagger_{p_1,p_2} & \tilde b^\dagger_{p_1,p_2}\cr}
\pmatrix{
0 & z\cr
z^* & 0\cr
}
\pmatrix{\tilde a_{p_1,p_2} \cr \tilde b_{p_1,p_2} \cr}
\end{equation}
where
\begin{equation}
\label{phase}
z= 1 +e^{- ip_1} +e^{+i p_2}.
\end{equation}
The three terms in this expression correspond to the three types of
bonds: horizontal, right leaning, and left leaning.

We see that due to the presence of two types of site, a spinor
structure $\psi=\pmatrix {a\cr b\cr}$ ``emerges'' naturally.  For a
given momentum, the Hamiltonian reduces to a two by two matrix
\begin{equation}
\label{hinp}
\tilde H(p_1,p_2)=K\pmatrix{
0 & z\cr
z^* & 0\cr
}.
\end{equation}
This is diagonalized with the eigenstates being two component spinors
\begin{equation}
\label{spinor}
\psi = {1\over \sqrt {2|z|}} \pmatrix{\sqrt z \cr \pm \sqrt {z^*}}.
\end{equation}
The corresponding eigenvalues are
\begin{equation}
\label{energy}
E(p_1,p_2)=\pm K|z|.
\end{equation}

The fermionic level energy vanishes when {$|z|$} does.  This occurs at
exactly two points
\begin{equation}
p_1=p_2=\pm 2\pi/3.
\end{equation}
Near these zeros, the energy behaves linearly in momentum.  This
linear behavior is exactly that of the 
relativistic Dirac equation.  

For our problem we want to consider excitations on the half filled
system.  The ``vacuum'' has all negative energy states filled.  This,
of course, has infinite negative energy, which should be subtracted to
find the energies of physical states.  And for rotations, we are free
to define things such that the vacuum state is invariant under such.

In this framework, one particle states are constructed by filling one
of the positive energy states.  Anti-particles correspond to holes in
the Dirac sea.  The concept of spin in two space dimensions is
somewhat different than in three.  In particular, we don't have
helicity states.  Rather we have rotations which are naturally thought
of as about an axis orthogonal to the spatial plane.  A state of
definite angular momentum $J$ transforms under a rotation by an angle
$\theta$ as
\begin{equation}
|\psi\rangle \longrightarrow e^{iJ\theta} |\psi\rangle.
\end{equation}
The basic idea here is to consider such a rotation on the vacuum with
one additional filled positive energy state.  We have half integer
spin if a rotation by $2\pi$ gives the wave function a minus sign.

\section{Topology and spin}
\label{topology}

\begin{figure}
\centering
\includegraphics[width=3in]{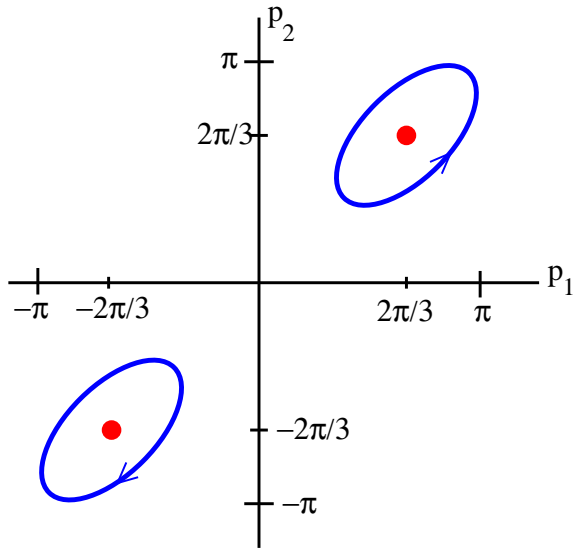}
\caption{When a curve of constant energy wraps around one of the Dirac
  points, the phase of the variable $z$ wraps around a unit circle and
  the spinor wave function acquires a minus sign.}
\label{contours}
\end{figure}

The way the spin arises in this model is related to a topological
behavior within the Brilloin zone.  Consider a contour of constant
energy near and wrapping around one of the zero points, as sketched in
Fig.~\ref{contours}.  As we traverse this contour, the phase of $z$ in
Eq.~(\ref{phase}) wraps nontrivially around the unit circle.  This
wrapping means one cannot collapse this contour without shrinking it
down to a Dirac point at $|z|=0$, a point where the energy vanishes.
Furthermore, as one fully goes around the contour, the spinor
expression in Eq.~(\ref{spinor}) becomes its negative.  This is the
behavior of a half integer spin system.  Indeed, the fermion spin has
``emerged.''  This is only possible because the momentum eigenstates
involved are non-local in terms of the underlying position space
operators.

In addition to spin, this model has an emergent chiral symmetry.
Since all hoppings couple $a$ and $b$ sites, the sign of the
Hamiltonian is reversed if we replace each $b$ operator by its
negative.  In particular, the Hamiltonian in Eq.~(\ref{hinp})
anti-commutes with the matrix
\begin{equation}
\sigma_3
=\pmatrix{1&0\cr 0&-1}.
\end{equation}

There is a close connection with the no-go theorem of Nielsen and
Ninomiya \cite{Nielsen:1980rz}.  Because of the periodicity of the
Brilloin zone, any contour expanded to the boundaries of this zone
cannot wrap $z$ non-trivially.  Given any Dirac cone, there must exist
another about which the topology unwraps.  This requires an even
number of Dirac cones.  Two such is the minimum possible without
breaking the symmetries.  And the chiral properties of the two cones
must be opposite.  Since the complex number $z$ rotates around the
origin in opposite directions, in this model the rotation gives
conjugate phases to the physical fermion states near the two cones.
One might think of the two cones as representing spin up and spin down
in the direction orthogonal to the spatial plane.

For the three dimensionional case discussed later we will see that we
get a separate four component spinor associated with each cone.  Each
then represents a Dirac fermion with both spin states present.  In
that case the two cones can be thought of as representing an isospin
symmetry, also emerging from the dynamics.

\section{Hopping in a magnetic field}
\label{2dmag}

We now turn to a rather different model that exhibits a similar
phenomenon to the graphene case.  Staying in two dimensions, consider
spin-less fermions hopping on a square lattice.  Apply to this system a
constant magnetic field transverse to the plane, giving phase factors
to the fermions as they pass around closed loops.  This problem was
studied extensively by D. Hofstadter in Ref.~\cite{Hofstadter:1976zz},
showing a rather complex structure depending sensitively on the
strength the applied field.  In particular, the spectrum with a
rational flux $p/q$ passing through each plaquette gives rise to $q$
bands.  The strength is measured in the natural units, where the flux
gives a phase of $e^{2\pi ip/q}$ in traversing around a fundamental
plaquette.

Here we concentrate on the case $p/q=1/2$; so there are two bands in
the spectrum.  This situation has some interesting mathematical
properties that have been frequently studied in the condensed matter
context; see for example
Refs.~\cite{PhysRevLett.73.2158,Wen:1990fv,Affleck:1987zz}.  We will
see that the two bands touch at two Dirac points, in direct analogy
with graphene.  We implement the magnetic field by placing a static
phase $Z_x(x,y)$ ($Z_y(x,y)$) representing a $U(1)$ gauge field on
each link in the positive $x$ ($y$) direction.  We label the sites by
the integer coordinates $(x,y)$.  The flux condition means that the
product of the link variables around any plaquette to be $-1$.  We
will see that this system is equivalent to what are known as staggered
fermions \cite{Kogut:1974ag,Susskind:1976jm,Sharatchandra:1981si}.

\begin{figure}
\centering
\includegraphics[width=2.5in]{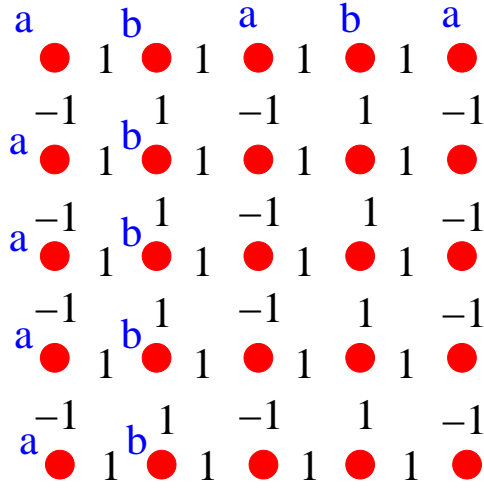}
\caption{Placing phases on the links of a two dimensional lattice so
  as to produce a static magnetic field of $1/2$ flux unit per plaquette.}
\label{maglat}
\end{figure}

The model leaves free a gauge freedom in selecting the link
variables.  Since the spectrum does not depend on the gauge choice, we are
free to pick a convenient one.  To proceed, we take 
\begin{equation}
Z_x(x,y)=1, \qquad Z_y(x,y)=(-1)^x.
\end{equation}
This arrangement of phases is sketched in Fig.~{\ref{maglat}}.  Since
this is just a gauge choice, the underlying physics retains the
underlying square symmetry.

With this choice we see that, as with graphene, there are two
types of site.  Let $(a^\dagger,a)$ label the fermion creation and
annihilation operators on the top of the negative $y$ bonds, and
correspondingly $(b^\dagger,b)$ for sites on top of positive $y$ bonds,
as labeled in the figure.

Again we find the spectrum of states via Fourier transform.  For this
gauge choice the translation symmetry in the $y$ direction remains
that of the underlying lattice.  On the other hand, the periodicity in
the $x$ direction involves a translation by twice the lattice spacing.
Thus at fixed momenta we find translationally invariant eigenstates
satisfy
\begin{equation}
\matrix{
&\psi(2x,y)=e^{2ip_xx+ip_yy}\ \psi(0,0)\cr
&\psi(2x+1,y)=e^{2ip_xx+ip_yy}\ \psi(1,0).
}
\end{equation}
In this way we wind up with the emergence of a
two component ``base spinor'' 
\begin{equation}
\Psi=\pmatrix  {\psi(0,0)\cr  \psi(1,0)}.
\end{equation}

Because of the modified translation symmetry, the Brilloin zone is
effectively ``half sized'' and we can take
\begin{equation}
\matrix{
&0\le p_y<2\pi\cr
&0\le p_x <\pi.\cr
}
\end{equation}
At fixed momentum we have a
two by two Hamiltonian matrix
\begin{equation}
H=K\pmatrix { 2\cos(p_y) & 1+e^{2ip_x}\cr
                    1+e^{-2ip_x} & 2\cos(p_y)\cr} 
\end{equation}
This is easily diagonalized to give
\begin{equation}
E=\pm 2K \sqrt{\cos^2(p_x)+\cos^2(p_y)}
\end{equation}
for the individual fermionic levels.  With this spectrum, the positive
and negative energy bands touch at just two points
$(p_x,p_y)=(\pi/2,\pi/2)$ and $(p_x,p_y)=(\pi/2,3\pi/2)$.  The system
displays two Dirac cones, just as in the graphene case.

As mentioned above, this is a rewriting of staggered fermions.  The
sign factors on the links arise from our gauge choice.  As this is
just a choice, the square symmetry of the underlying lattice is
preserved.

\section{Higher dimensions}
\label{3dstag}

This procedure immediately generalizes to three spatial dimensions.
Thread a half unit of magnetic flux through every plaquette of a cubic
lattice.  As before, it is useful to select a convenient gauge for the
phases $Z$ on the lattice bonds.  For this we take 
\begin{equation}
\matrix{
Z_x(x,y,z)=1\cr
Z_y(x,y,z)=(-1)^x\cr
Z_z(x,y,z)=(-1)^{x+y}\cr
}
\end{equation}
where we label the sites by the integer coordinates $(x,y,z)$.
Translation invariance is by one site in the $z$ direction but by two
in the $x$ and $y$ directions.  
Use translation invariance look for a solution of form
\begin{equation}
\matrix{
\psi(2x,2y,z)=\exp(2ip_x x+2ip_yy+ip_z z)\psi(0,0,0)\cr
\psi(2x+1,2y,z)=\exp(ip_x (2x+1)+2ip_yy+ip_z z)\psi(1,0,0)\cr
\psi(2x,2y+1,z)=\exp(2ip_x x+ip_y(2y+1)+ip_z z)\psi(0,1,0)\cr
\psi(2x+1,2y+1,z)=\exp(ip_x (2x+1)+ip_y(2y+1)+ip_zz)\psi(1,1,0).\cr
}
\end{equation}
This translates any fixed momentum wave function down to four base
components
\begin{equation}
\pmatrix{\psi(0,0,0)\cr
\psi(1,0,0)\cr
\psi(0,1,0)\cr
\psi(1,1,0)\cr
}.
\end{equation}
A four component spinor structure emerges.

The complete Brilloin zone is covered by $0\le p_x < 2\pi, 0\le p_{y,z}<\pi$,
i.e. one quarter of the naive zone.  
This gives a four by four reduced Hamiltonian
\begin{equation}
H=16\pmatrix{
\cos(p_z) & \cos(p_x) & \cos(p_y) & 0  \cr
\cos(p_x) & -\cos(p_z) & 0 & -\cos(p_y)\cr 
\cos(p_y) & 0 & -\cos(p_z) & \cos(p_x) \cr
0 & -\cos(p_y) & \cos(p_x) & \cos(p_z) \cr
}.
\end{equation}
This can be written more compactly as
\begin{equation}
H=16(\cos(p_x) I\otimes\sigma_1 + \cos(p_y) \sigma_1\otimes \sigma_3
+\cos(p_z) \sigma_3\otimes \sigma_3).
\end{equation}
The three terms anti-commute; so, energy
eigenvalues are
\begin{equation}
E=\pm 16\sqrt{\cos^2(p_x)+\cos^2(p_y)+\cos^2(p_z)}.
\end{equation}
In the restricted Brilloin zone, we again have two Dirac cones, one at
$\vec p=(\pi/2,\pi/2,\pi/2)$ and the other at $\vec
p=(3\pi/2,\pi/2,\pi/2)$.  We have a Hamiltonian version of staggered
fermions.  In three dimensions we have only two doublers, the minimum
required if there is to be a chiral symmetry.

For connection with the more usual continuum Hamiltonian, identify
$i\gamma_0\vec\gamma=(I\otimes\sigma_1,\sigma_1\otimes \sigma_3,
\sigma_3\otimes \sigma_3)$.  Given that $\gamma_0$ should anti-commute
with these and $\gamma_5$ commute, we are led to the gamma matrix
convention
\begin{equation}
\matrix{
\gamma_1=\sigma_2\otimes\sigma_2\cr
\gamma_2=\sigma_3\otimes I\cr
\gamma_3=\sigma_1 \otimes I \cr
\gamma_0=\sigma_2\otimes \sigma_3\cr
\gamma_5=\sigma_2\otimes \sigma_1.\cr
}.
\end{equation}

Since the Hamiltonian only contains nearest neighbor hoppings, its
sign would be changed by changing the signs on the creation and
annihilation operators on all sites of a given parity.  Thus it
anti-commutes with $(-1)^{x+y+z}$ which in the continuum should be
represented by $\gamma_0\gamma_5$.

Near the zeros in the Brilloin zones, the Dirac cones display a linear
dispersion.  These come with opposite effective chirality.  Thus the
theory which has emerged has two ``flavors'' with opposite chirality.
The physical chiral symmetry is actually a ``flavored'' symmetry.
This is consistent with the anomaly, which forbids a flavor
singlet chiral symmetry. 

\section{Path integrals}
\label{paths}

So far the discussion has been in terms of a hopping Hamiltonian.
Usually lattice gauge theory is discussed in terms of the Euclidean
space path integral including a Dirac operator appearing in a
quadratic form $\overline\psi D\psi$.  Here $\overline\psi$ and $\psi$
are independent Grassmann fields.  In a continuum discussion $D$ is
formally an anti-Hermitean operator, although this is not generally
true on the lattice.

The procedure in the previous section of inserting phase factors on
the lattice links generalizes immediately to a four dimensional
lattice, and in this way we can find a corresponding Hermitean
Hamiltonian $H$.  One can try to use this form directly in a path
integral, using $D=iH$.  With a corresponding gauge choice, we have
peridicity by two lattice sites in three of the dimensions and
periodicity of one in the remaining.  In this way the basic spinor
that emerges has 8, rather than the 4 components of the usual Dirc
theory.  In terms of the relativistic fermions that appear, there is
an extra doubling.  This gives rise to a total of four ``tastes,'' a
famous aspect of staggered fermions.  It is possible to break the
degeneracy of these states by adding Hermitean terms to $D$ that break
the chiral symmetry \cite{Adams:2010gx,Hoelbling:2010jw}.  That
procedure is similar in spirit to the Wilson fermion construction.

\blockcomment

But since we have a three dimensional Hamiltonian formulation with
only two doublers, it is natural to ask if this can lead to a
minimally doubled staggered action for the path integral.  This would
be particularly convenient because the physical world has two light
quark species, rather the four that are natural in the usual staggered
approach.

This procedure is possible, but one must use care to properly treat
the connection between the Hamiltonian and path integral approaches.
One method is to use an old exact relationship \cite{Creutz:1999zy}
for the trace of a a product of $n$ normal ordered operators in a
Hilbert space generated by fermionic creation and annihilation
operators to a path integral over a set of Grassmann variables
\begin{eqnarray}
& {\rm Tr} \left( :f_1(a^\dagger,a)::f_2(a^\dagger,a):
\ldots :f_n(a^\dagger,a): \right)\cr
&=\int (d\psi d\psi^*)
f_1(\psi_1^*,\psi_1)\ldots f_n(\psi_n^*,\psi_n)
\ e^{\sum_j \psi_j^*(\psi_j-\psi_{j-1})}.
\end{eqnarray}
Here $\psi_n$ and $\psi_n^*$ are {independent} Grassmann variables.
In this formula one should identify $\psi_0=-\psi_n$; i.e. one should
use anti-periodic boundaries.  It is important to note the appearance
of an asymmetric discrete derivative \cite{Creutz:1999zy} in what will
become the time direction.  The Dirac operator we will derive has both
Hermitean and anti-Hermitean parts.

To obtain a fermion action for a path integral, divide time into $N_t$
slices and apply this formula to
\begin{equation}
{\rm Tr} \left(\left (:e^{-\beta H/N_t}:\right)^{N_t}\right).
\end{equation}
From this we find an action $S$ that satisfies
\begin{equation}
{\rm Tr}e^{-\beta H} \sim \int (d\psi d\psi^*)e^{-S(\psi^*\psi)}.
\end{equation}
Leaving out the gauge field factors for simplicity, the free action
can be written
\begin{equation}
S=\sum_t\sum_{j}
\left(
\psi^*_{j,t}(\psi_{j,t}-\psi_{j,t-1})
+\sum_{k=1}^3 
{\beta Z(j,k)\over N_t}\left(\psi^*_{j+e_k,t} \psi_{j,t}
+\psi^*_{j,t} \psi_{j+e_k,t}\right)
\right)
\end{equation}
with the $Z$'s being the staggered phase factors on the spatial links.  
This gives a four dimensional staggered action which is
minimally doubled with two ``tastes.''  Again, this is the
minimal number required for a chiral symmetry to survive.

This action is not what is usually used and has some interesting
features.  First, it contains both Hermitean and anti-Hermitean parts.
In this it shows a certain similarity to Wilson fermions
\cite{Wilson:1975id}.  Second, the time direction is treated
differently than the spatial coordinates.  The presence of a
non-symmetric hopping in this direction explicitly breaks $4d$
hyper-cubic symmetry.  The breaking of hyper-cubic symmetry indicates
that on adding the gauge fields one should expect to require
additional counter-terms to maintain the proper Lorentz symmetry.  In
particular, the ``speed of light'' for both the fermions and the gauge
fields can be renormalized \cite{Capitani:2010ht}.  Third, when gauge
fields are present, in general the corresponding fermion determinant
need not be positive.  This introduces a potential ``sign problem''
into any Monte Carlo treatment.  It is unclear how severe this problem
would be in practice.

\endcomment

A minimally doubled formulation of staggered fermions with only two
``tastes'' appears in \cite{vandenDoel:1983mf,Golterman:1984cy}.  This
form preserves the hyper-cubic symmetry, but suffers from a
non-positive determinant.  There are a variety of other known 4d
minimally doubled chiral formulations for fermions
\cite{Creutz:2010cz} that do not suffer from a sign problem.  All,
however, appear to break hyper-cubic symmetry in some way.  This in
general introduces new renormalization counterterms; in particular,
the ``speed of light'' for both the fermions and the gauge fields can
be renormalized \cite{Capitani:2010ht}.  Whether this is always
necessary for a four dimensional minimally doubled and chiral
discretization is not known.

Since we have a three dimensional Hamiltonian formulation with
only two doublers, it is natural to ask if this can lead to a
minimally doubled staggered action for the path integral.  This would
be particularly convenient because the physical world has two light
quark species, rather the four that are natural in the usual staggered
approach.

This procedure is possible, but one must use care to properly treat
the connection between the Hamiltonian and path integral approaches.
One method is to use an old exact relationship \cite{Creutz:1999zy}
for the trace of a a product of $n$ normal ordered operators in a
Hilbert space generated by fermionic creation and annihilation
operators to a path integral over a set of Grassmann variables.

The resulting action is not what is usually used for staggered
fermions and has some interesting features.  First, it contains both
Hermitean and anti-Hermitean parts.  In this it shows a certain
similarity to Wilson fermions \cite{Wilson:1975id}.  Second, the time
direction is treated differently than the spatial coordinates.  The
presence of a non-symmetric hopping in this direction explicitly
breaks $4d$ hyper-cubic symmetry.  The breaking of hyper-cubic
symmetry indicates that on adding the gauge fields one should expect
to require additional counter-terms to maintain the proper Lorentz
symmetry.  Third, when gauge fields are present, the corresponding
fermion determinant need not be positive.  This introduces a potential
``sign problem'' into any Monte Carlo treatment.  It is unclear how
severe this problem would be in practice.  And fourth, the explicit
chiral symmetry of the Hamiltonian becomes hidden and it is unclear
whether lattice artifacts from the discretization in the time
direction would introduce the need for an additional mass counterterm.

\section{Including gauge fields}
\label{gauge}

The above discussion has been in terms of free fermions.  Nonetheless,
it is straightforward to insert gauge degrees of freedom.  If we
consider an $SU(N)$ gauge group, traditionally this would be done by
inserting group matrices on the lattice links.  These are independent
of the fixed phases $Z$ on the links.  Since the latter are all plus
or minus one, we might as well consider them as an auxiliary $Z_2$
gauge field giving a factor of $Z_{ij}$ on the link from site $i$ to
site $j$.  When a fermion hops it picks up the product $U_{ij}Z_{ij}$
with $U_{ij}$ being the link variable from the gauge group.

One can think of the theory as having two gauge couplings, a
$\beta\sim {1\over g^2}$ for the non-Abelian $SU(N)$ and a second
$\beta_z$ for the $Z_2$ factors.  The above model arises in taking the
limit $\beta_z\rightarrow -\infty$, which drives all $Z_2$ plaquettes
to $-1$.  As mentioned above, this limit is exactly staggered
Hamiltonian lattice gauge theory

A rather interesting special case occurs when the color group is
even. Although this is not relevant to the physical $SU(3)$ of color,
it does apply to the simple $SU(2)$ gauge group.  With an even number
of colors, then the phase factor $-1$ is in the gauge group.  In the
link product $U_{ij}Z_{ij}$, one can then absorb the $Z$ factor into
the $U$ matrices.  The $SU(N)$ measure is invariant under this.  After
such an absorption, the only remnant of the $Z_2$ factors is that the
coefficient $\beta$ for the $SU(N)$ plaquette term changes sign.  We
conclude that for any even $N$, a conventional gauge theory of
spin-less fermions at negative $\beta$ is equivalent to staggered
fermions.  This does not work for $SU(3)$ because $-1$ is not an
element of the gauge group.

\section{Gauge fields and topology}
\label{gaugetopology}

In this section we change subject slightly and discuss the connections
between lattice fermion fields and the topology of gauge fields.  In
a continuum field theory there is a well known index theorem
\cite{Atiyah:1963zz}, where
\begin{equation}
\nu=n_+-n_-.
\end{equation}
Here $n_\pm$ counts the number of zero modes of the continuum Dirac
operator of $\pm$ chirality.  The other side of the equation,
$\nu$ represents the topological index of the gauge field.

This relation represents the heart of the anomaly.  If the numbers of
zero modes of each chirality are unequal, formally ${\rm
  Tr}\ \gamma_5=\nu$.  This come about because all non-zero
eigenvalues of the Dirac operator come in complex conjugate pairs that
cancel in this trace.  The main consequence is that the naive chiral rotation
\begin{equation}
\psi \rightarrow e^{i\gamma_5 \theta} \psi
\end{equation}
changes the integration measure in the path integral \cite{Fujikawa:1979ay}
\begin{equation}
d\psi\rightarrow 
e^{i\theta {\rm Tr}\gamma_5}\ d\psi=
e^{i\nu\theta}\ d\psi.
\end{equation}
Because of this, a chiral rotation of the mass term
\begin{equation}
m\ \overline\psi\psi \rightarrow m\ \overline\psi
e^{i\gamma_5\theta}\psi
\end{equation}
results in an inequivalent theory.  This is the famous ``theta
vacuum,'' which violates CP symmetry when the rotation is non-trivial.

While all this is standard in a continuum formulation, on the lattice
it becomes a bit trickier.  With the cutoff in place $\gamma_5$ is
always a finite traceless matrix, in apparent conflict with the above.
To emphasize this point, consider any lattice Dirac operator $D$.  For
simplicity assume this satisfies gamma five hermiticity
\begin{equation}
\gamma_5 D \gamma_5=D^\dagger.
\end{equation}
All the fermion operators used in practice in lattice gauge theory
satisfy this (except for twisted mass, which brings in an additional
isospin rotation).

Now consider dividing $D$ into Hermitean and anti-Hermitean parts
$D=K+M$
\begin{eqnarray}
&K=(D-D^\dagger)/2,\cr
&M=(D+D^\dagger)/2.
\end{eqnarray}
These automatically satisfy
\begin{eqnarray}
[K,\gamma_5]_+=0,\cr
[M,\gamma_5]_-=0.
\end{eqnarray}
With these definitions, 
\begin{equation} 
M\rightarrow e^{i\theta\gamma_5}M
\end{equation}
is automatically an exact symmetry of the
determinant  
\begin{equation}
|K+M|=| e^{i\theta\gamma_5/2}(K+M) e^{i\theta\gamma_5/2}|
=|K+e^{i\theta\gamma_5}M|.
\end{equation}
This seems to contradict the continuum discussion.  Indeed, where is
the anomaly?  In the above models, the answer lies with the doublers.
Half of them use $\gamma_5$ and half $-\gamma_5$ for chiral rotations.
In this case the naive chiral symmetry is actually flavored, and there
is no contradiction.

Note that the above discussion was completely general and applies even
to Wilson fermions, which are normally thought of as breaking chiral
symmetry.  What happens in this case depends crucially on the
doublers, which have been given masses of order the cutoff.  The above
rotation $M\rightarrow e^{i\theta\gamma_5}M$ also rotates their
phases.  The important point is that the physical $\Theta$ is a
relative angle arising when one independently rotates the fermion mass
and the Wilson term; this was pointed out long ago by Seiler and
Stamatescu \cite{Seiler:1981jf}. In a sense, we still have a flavored
chiral symmetry.

This interpretation also applies to the overlap operator
\cite{Narayanan:1993sk}.  In this case the eigenvalues of the Dirac
operator lie on a complex circle.  For every zero eigenmode there
exists a heavy counterpart on the opposite side of the circle.  The
above rotation of Hermitean part rotates both the low mode and the
heavy mode as well.  Formally the anomaly brings in another chiral
matrix $\hat\gamma_5$ defined so that $D\gamma_5=-\hat\gamma_5 D$.  In
this approach the index theorem becomes
\begin{equation} 
\nu={\rm Tr} (\gamma_5+\hat\gamma_5)/2
\end{equation}
and need not vanish when $\hat\gamma_5$ is not traceless.

This discussion reveals an interesting message for continuum QCD.  The
physical parameter $\Theta$ can be moved around and placed on the mass
term for any one flavor at will.  In particular, $\Theta$ can be
entirely moved into the phase of the top quark mass.  This has the
non-intuitive consequence that the top quark properties are relevant
to the low energy physics of QCD.  The traditional decoupling theorems
\cite {Appelquist:400372} don't apply non-perturbatively when the
masses have phases.

\section{Conclusions}
\label{conclusions}

We have discussed a variety of models for spinless fermions hopping on
simple lattices for which the excitations on the Dirac sea can carry
spin.  This is required by the relativistic form of spectrum.  The
phenomenon has close connections with chiral symmetry and the
topological protection from additive mass renormalization.  For the
phenomenon to occur, doublers are required to appear.  And the entire
picture is intimately entwined with the the possibility of a CP
violating parameter $\Theta$ in QCD.

%\bibliographystyle{unsrt}
%\bibliographystyle{aps}
%\bibliography{../../papers/references,../../papers/creutz}

\end{document}